\documentclass[runningheads,a4paper]{llncs}

\usepackage[caption=false]{subfig}
\usepackage{amssymb}
\usepackage{graphicx}
\usepackage{algorithmic}
\usepackage{algorithm}
\usepackage{float}
\usepackage{amsmath}
\newfloat{algorithm}{t}{lop}

\usepackage[utf8]{inputenc}

\usepackage{url}
\newcommand{\keywords}[1]{\par\addvspace\baselineskip
\noindent\keywordname\enspace\ignorespaces#1}

\include{Qcircuit}

\begin{document}

\mainmatter  

\title{Circular CNOT Circuits: Definition, Analysis and Application to Fault-Tolerant Quantum Circuits}

\titlerunning{Circular CNOT Circuits}

\author{Alexandru Paler$^*$}
\authorrunning{Circular CNOT Circuits}

\institute{$^*$Facult. de Matematică şi Informatică, Universitatea Transilvania, Braşov, Romania}

\toctitle{Lecture Notes in Computer Science}
\tocauthor{Authors' Instructions}
\maketitle

\begin{abstract}
The work proposes an extension of the quantum circuit formalism where qubits (wires) are circular instead of linear. The left-to-right interpretation of a quantum circuit is replaced by a circular representation which allows to select the starting point and the direction in which gates are executed. The representation supports all the circuits obtained after computing cyclic permutations of an initial quantum gate list. Two circuits, where one has a gate list which is a cyclic permutation of the other, will implement different functions. The main question appears in the context of scalable quantum computing, where multiple subcircuits are used for the construction of a larger fault-tolerant one: can the same circular representation be used by multiple subcircuits? The circular circuits defined and analysed in this work consist only of CNOT gates. These are sufficient for constructing computationally universal, fault-tolerant circuits formed entirely of qubit initialisation, CNOT gates and qubit measurements. The main result of modelling circular CNOT circuits is that a derived Boolean representation allows to define a set of equations for $X$ and $Z$ stabiliser transformations. Through a well defined set of steps it is possible to reduce the initial equations to a set of stabiliser transformations given a series of cuts through the circular circuit.
\end{abstract}

\section{Motivation}

The quantum circuit formalism is a generally accepted representation of quantum information processing. It is mainly inspired by the classical circuit representation, where input information is transformed through the application of gate sequences into output information. The main differences between classical and quantum circuits are that the latter have an equal number of inputs and outputs, do not accept FANIN or FANOUT and the quantum gates  represent reversible transformations required by the unitarity of quantum mechanics, unlike classical gates (e.g. the classical AND gate) which are not reversible.

\begin{figure}[t!]
\centering
\subfloat[ ]
{
\Qcircuit @C=.3em @R=.35em {
		& \ctrl{1} & \ctrl{2} & \qw \\
		& \targ & \qw & \qw \\
		& \qw & \targ & \qw 
	}
}
\hfil
\subfloat[ ]
{
\Qcircuit @C=.3em @R=.35em {
		& \ctrl{2} & \ctrl{1} & \qw \\
		& \qw & \targ & \qw  \\
		& \targ & \qw & \qw
	}
}
\hfil
\subfloat[ ]
{
\Qcircuit @C=.3em @R=.35em {
		& \ctrl{2} & \qw & \qw \\
		& \qw & \ctrl{1} & \qw \\
		& \targ & \targ & \qw 
	}
}
\hfil
\subfloat[ ]
{
\Qcircuit @C=.3em @R=.35em {
		& \qw & \ctrl{2} & \qw \\
		& \ctrl{1} & \qw & \qw \\
		& \targ & \targ & \qw 
	}
}
\hfil
\subfloat[ ]
{
\label{circ:swap}
\Qcircuit @C=.3em @R=.35em {
		& \ctrl{1} & \targ & \ctrl{1} & \qw \\
		& \targ & \ctrl{-1} & \targ & \qw \\
	}
}
\hfil
\subfloat[ ]
{
\label{circ:swapperm}
\Qcircuit @C=.3em @R=.35em {
		& \ctrl{1} & \ctrl{1} & \targ & \qw \\
		& \targ & \targ & \ctrl{-1} & \qw
	}
}
\caption{CNOT circuits: a-d) CNOT commutativity rules; e) the SWAP circuit; f) a cyclic permutation of the SWAP circuit.}
\label{circ:cnotcomm}
\end{figure}
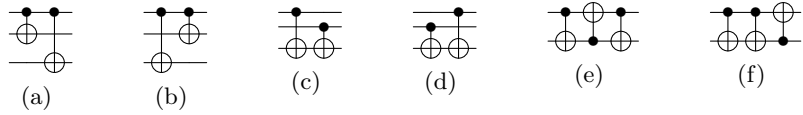

A quantum circuit is specified as a gate sequence containing gates from an universal gate set, and in the context of practical quantum computing this set is $\{CNOT, T, P, V\}$. $T$ and $P$ are $\pi/4$ and $\pi/2$ rotations around the $Z$-axis and $V$ is the $\pi/2$ rotation around the $X$-axis of the Bloch sphere \cite{NC00}. These gates are sufficient for approximating any quantum computation with arbitrary precision, and are preferred because they have known fault-tolerant implementations used within error corrected quantum computing architectures \cite{bravyi2005universal}. The gate sequence introduces a temporal ordering of information processing, although this ordering is not entirely strict because some gates can commute (Fig.~\ref{circ:cnotcomm}).

Universal fault-tolerant quantum circuits can be represented as ICM circuits which are formed entirely of qubit (I)nitialisations, (C)NOT gates and qubit (M)easurements \cite{paler2015fully}. The circuits include only CNOT gates, because rotational gates are implemented by teleportation mechanisms \cite{NC00,paler2015fully} and rotations are achieved by initialising certain qubits in specific ancillary states. The computational universality of ICM circuits is based on the choice of the qubit initialisation and measurement basis: the $Y$ and the $A$ basis can be chosen in addition to the $X$ and $Z$ basis \cite{bravyi2005universal}. Therefore, ICM circuit qubits can be initialised into $\ket{0}$, $\ket{+}$, $\ket{Y}=\ket{0}+i\ket{1}$ and $\ket{A}=\ket{0}+e^{\pi/4}\ket{1}$, and can be measured in the $X,Y,Z,A$ basis \cite{FMM13}. The $\ket{Y}$ and $\ket{A}$ states are required for implementing the $T$ (Fig.~\ref{circ:tgate}), $P$ (Fig.~\ref{circ:pgate}) and $V$ (Fig.~\ref{circ:vgate}) gate.

\begin{figure}[t!]
\centering
\subfloat[ ]{
	\label{fig:swap0}
	\includegraphics[width=0.18\columnwidth]{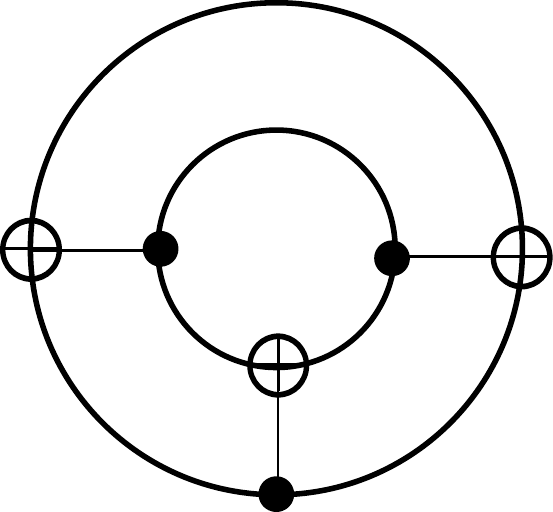}
}
\hfil
\subfloat[ ]{
	\label{fig:swap1}
	\includegraphics[width=0.18\columnwidth]{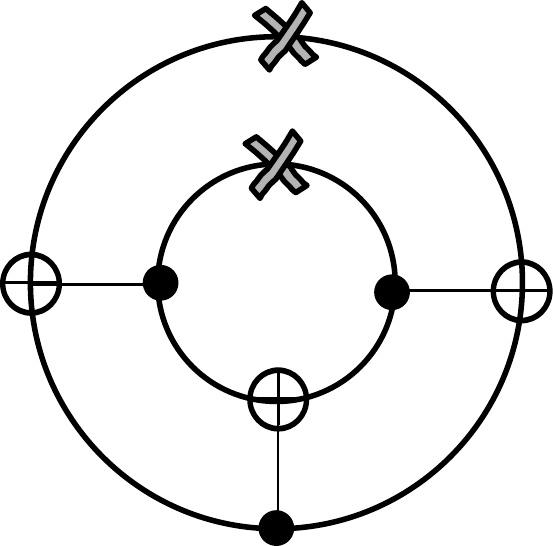}
}
\hfil
\subfloat[ ]{
	\label{fig:swap2}
	\includegraphics[width=0.18\columnwidth]{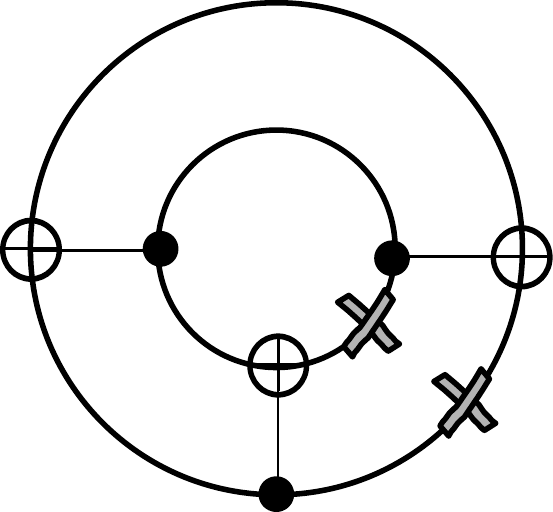}
}
\hfil
\subfloat[ ]{
	\label{fig:swap3}
	\includegraphics[width=0.18\columnwidth]{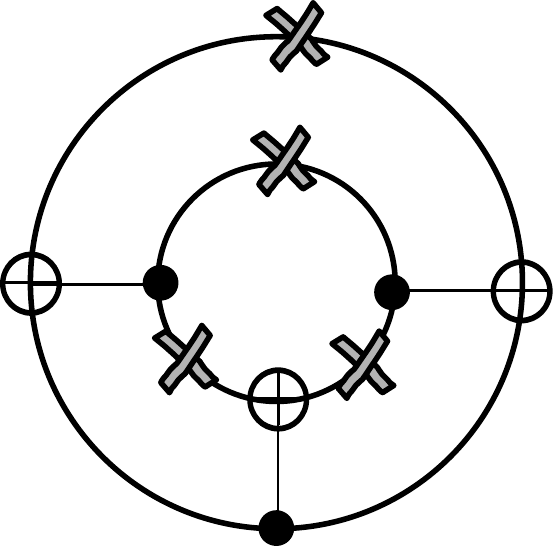}
}
\hfil
\subfloat[ ]{
	\label{fig:swap4}
	\includegraphics[width=0.18\columnwidth]{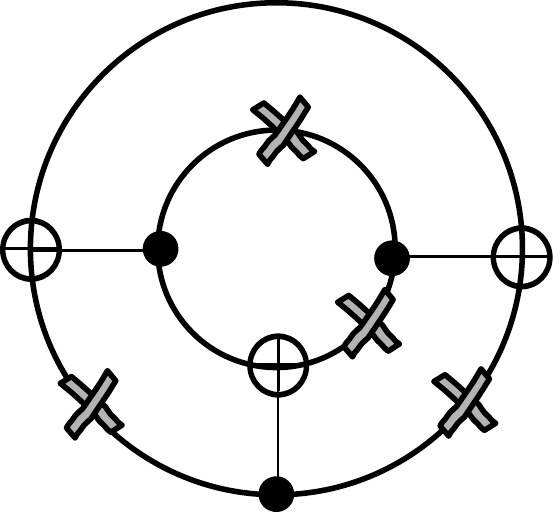}
}
\caption{Cutting a circular CNOT circuit results in ICM circuits: a) the initial circular SWAP circuit; b)  radial cut for the linear SWAP circuit; c) radial cut for the linear CNOT circuit; d) cuts for the teleported CNOT circuit; (Fig.~\ref{circ:rcnot}) e) cuts for the selective destination teleportation circuit (Fig.~\ref{circ:seldest}).}
\end{figure}

Multiple circuits share the same CNOT gates circuit structure resulting after not considering the initialisations and measurements of an arbitrary ICM circuit. This is illustrated by the example of the SWAP circuit (Fig.~\ref{circ:swap}). The circuit has two qubit lines and three CNOT gates. Consider that, without being offered any definition, the circular representation from Fig.~\ref{fig:swap0} results after joining the inputs and the outputs. The initial SWAP circuit can be reconstructed after making a \emph{cut} on each of the circular qubits, so that there is no case where two CNOTs have the same control or target. However, if the cuts are made as indicated in Fig.~\ref{fig:swap2}, the result will be a circuit that implements a single CNOT, because the other two cancel out (Fig.~\ref{circ:swapperm}).

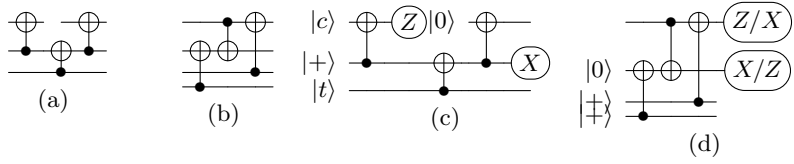
\begin{figure}[t!]
\centering
\subfloat[ ]{
\label{circ:rcnot}
\Qcircuit @C=.3em @R=.35em {
		& \targ & \qw & & \targ & \qw \\
		& \ctrl{-1} & \qw & \targ & \ctrl{-1} & \qw\\
		& \qw & \qw & \ctrl{-1} & \qw & \qw
}
}
\hfil
\subfloat[ ]{
\label{circ:seldest}
\Qcircuit @C=.3em @R=.35em {
		& \qw & \ctrl{1} & \targ & \qw \\
		& \targ & \targ & \qw & \qw\\
		& \qw & \qw & \ctrl{-2} & \qw\\
		& \ctrl{-2} & \qw & \qw & \qw
	}
}
\hfil
\subfloat[ ]{
\label{circ:icmrcnot}
\Qcircuit @C=.3em @R=.35em {
		\lstick{\ket{c}}& \targ & \qw & \measure{Z} & & & \lstick{\ket{0}} & \targ & \qw & \qw \\
		\lstick{\ket{+}}& \ctrl{-1} & \qw &\qw & \targ & \qw & \qw & \ctrl{-1} & \qw & \measure{X}\\
		\lstick{\ket{t}}& \qw & \qw & \qw & \ctrl{-1} & \qw & \qw & \qw & \qw & \qw
}
}
\hfil
\subfloat[ ]{
\label{circ:icmseldest}
\Qcircuit @C=.3em @R=.35em {
		& \qw & \ctrl{1} & \targ & \qw & \measure{Z/X}\\
		\lstick{\ket{0}} & \targ & \targ & \qw & \qw& \measure{X/Z}\\
		\lstick{\ket{+}} & \qw & \qw & \ctrl{-2} & \qw\\
		\lstick{\ket{+}} & \ctrl{-2} & \qw & \qw & \qw
	}
}
\caption{Circuits after cutting the circular SWAP circuit: a) as in Fig.~\ref{fig:swap3}; b) as in Fig.~\ref{fig:swap4}. The ICM versions of the previous two CNOT structures is obtained after choosing appropriate qubit initialisation and measurements basis: c) remote CNOT circuit; d) selective destination teleportation, where the measurement of the two upper qubits dictates on which qubit (third or fourth) the first qubit is to be teleported. In general, the $\ket{0}$ state can be replaced with an arbitrary state.}
\end{figure}

The circular representation of the SWAP can be cut in different ways, and the resulting circuits will have different functionality. The circuit from Fig.~\ref{circ:rcnot} is obtained by executing the cuts from Fig.~\ref{fig:swap3}. Furthermore, if the cyclic permutation of SWAP is cut according to Fig.~\ref{fig:swap4}, the resulting circuit will be the one from Fig.~\ref{circ:seldest}. By augmenting both resulting circuits with specific qubit initialisation and measurement bases, these have practical functional interpretations: Fig.~\ref{circ:rcnot} depicts a remote CNOT (Fig.~\ref{circ:icmrcnot}), and Fig.~\ref{circ:seldest} implements the selective destination teleportation circuit \cite{fowler2012time} (Fig.~\ref{circ:icmseldest}).

\section{Circular CNOT Circuits}

A circular CNOT circuit was presented in the previous section without any definition or discussing its properties. In the following paragraphs definitions will be introduced and explained. It should be noted that the notion of treating a circuit in a circular fashion is the basis for the approach to template matching \cite{maslov2003simplification}, where templates are considered cyclic gate sequences. In contrast, the circuits presented in this paper have circular wires that can be cut. This leads to a set of implementable fault-tolerant quantum circuits requiring different amount of qubits.

\begin{definition}
A circular CNOT circuit has circular qubit wires and consists entirely of CNOT gates, thus it has no inputs or outputs.
\end{definition}

The circular CNOT circuits proposed herein are not able to process information because of to their lack of inputs and outputs, but can be transformed into linear quantum circuits by cutting the circular wires. Quantum circuit reversibility is captured by the circular wire representation, and the temporal ordering of the gates is dictated by the direction in which the wires are traversed. Therefore, after cutting the wires, depending on the direction chosen, some wire end points are the inputs and others represent outputs. 

\begin{definition}
A cut is an interruption of a circular qubit wire that generates two end points associated to an input or an output.
\end{definition}

A set of cuts is \emph{correct} if it does not lead to CNOTs intersecting themselves in the resulting circuit. 
It can be shown that at least one \emph{radial cut} across all the wires is necessary for generating a valid quantum circuit: each cut introduces two end points; if two cuts generate end points which are not co-linear on the same radius then, after choosing any traversal direction, at least one CNOT will have one of its control/target after an input and right before an output.

\begin{definition}
A linear quantum circuit is the result of performing two operations: 1) cut correctly at least once each circular wire of a circular CNOT circuit; 2) chose a direction in which to traverse the CNOTs (clockwise, counter-clockwise).
\end{definition}

\section{Boolean Model of Circular CNOT Circuits}

Classical circuits can be modelled using Boolean formula, and this section shows that circular CNOT circuits have a Boolean representation, too. This is not surprising as the CNOT gate is a reversible gate. However, the Boolean model uses the fact that the CNOT gate is a stabiliser gate \cite{aaronson2004improved,NC00} whose transformations have a Boolean representation. An exact definition of the introduced Boolean variables is offered only after discussing the effect of the cuts on the circular representation.

\subsection{Stabiliser Transformations}

The Pauli matrices $I,X,Y,Z$ play a central role in the definition of quantum circuits. In the following the discussion will focus on $X$ and $Z$, because $Y=iXZ$ and $I$ is the $2\times 2$ identity matrix. The matrices can be decomposed into $\pm 1$ eigenvalues with corresponding eigenvectors. The eigenvectors of $Z$ are $\ket{0}$ and $\ket{1}$, and the ones of $X$ are $\ket{+}=\frac{1}{\sqrt{2}}(\ket{0}+\ket{1})$ and $\ket{-}=\frac{1}{\sqrt{2}}(\ket{0}-\ket{1})$. The states $\ket{0}$ and $\ket{+}$ are $+1$ eigenvectors, and $\ket{1},\ket{-}$ are $-1$ eigenvectors respectively.
\begin{eqnarray*}
\small{
\begin{array}{cccccccccccc}
I & = &
\left( \begin{array}{cc}
	1 & 0\\
	0 & 1
	\end{array} \right)
&
 Y & = &
\left( \begin{array}{cc}
	0 & -i\\
	i & 0
	\end{array} \right)
&
X & = &
\left( \begin{array}{cc}
	0 & 1\\
	1 & 0
	\end{array} \right)
&
 Z & = &
\left( \begin{array}{cc}
	1 & 0\\
	0 & -1
	\end{array} \right)
\end{array}}
\end{eqnarray*}

An operator $M$, consisting of $N$ tensor products of Pauli operators, stabilises the $N$-qubit state $\ket{q}$, if $\ket{q}$ is a $+1$ eigenvector of $M$. Therefore, for example, $X$ stabilises $\ket{+}$ and $-Z$ stabilises $\ket{1}$. The matrix $I$ stabilises any state. The action of certain quantum gates (Clifford gates), which includes CNOT, can be formulated entirely in terms of stabiliser transformations. The following equations illustrate how the input states of a CNOT ($_c$ denotes the control and $_t$ the target) are transformed. For example, Eq.~\ref{eq:xi}, shows that if the control qubit is stabilised by $X$ (is in the $\ket{+}$ state), after the CNOT both the control and the target are stabilised by $X$.  The set of four stabiliser transformations below are a complete description of the function of a CNOT.  
\begin{align}
X_cI_t &\rightarrow X_cX_t \label{eq:xi} \\
I_cX_t &\rightarrow I_cX_t \label{eq:ix} \\
Z_cI_t &\rightarrow Z_cI_t \label{eq:zi} \\
I_cZ_t &\rightarrow Z_cZ_t \label{eq:iz}
\end{align}

\subsection{A Single CNOT}
\label{sec:cnot}

\begin{figure}[t!]
\centering
\subfloat[ ]{
	\label{fig:cnotxvars}
	\includegraphics[width=0.12\columnwidth]{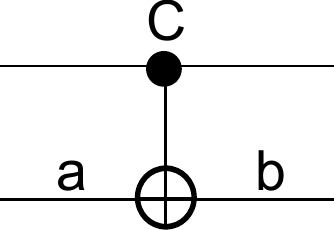}
}
\hfil
\subfloat[ ]{
	\label{fig:cnotzvars}
	\includegraphics[width=0.12\columnwidth]{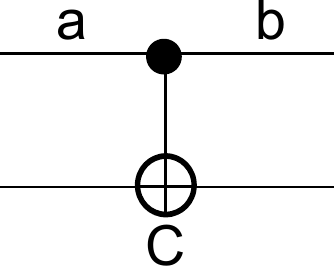}
}
\hfil
\subfloat[ ]{
	\label{fig:cnotvars}
	\includegraphics[width=0.12\columnwidth]{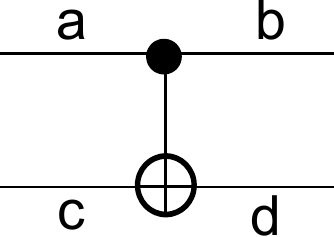}
}
\hfil
\subfloat[ ]{
	\label{fig:swapxvars}
	\includegraphics[width=0.18\columnwidth]{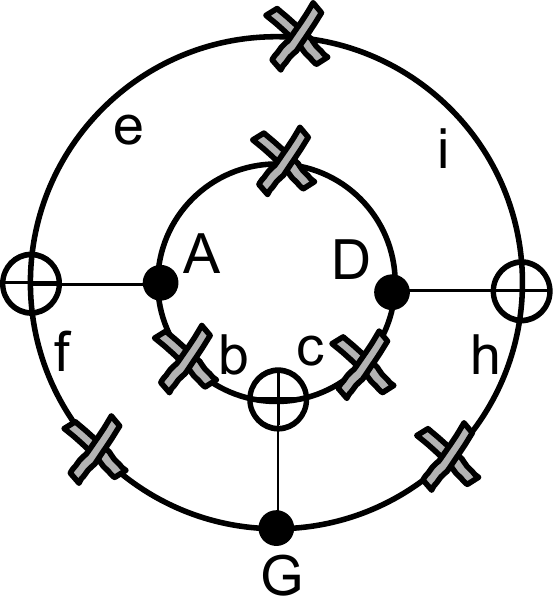}
}
\hfil
\subfloat[ ]{
	\label{fig:swapzvars}
	\includegraphics[width=0.18\columnwidth]{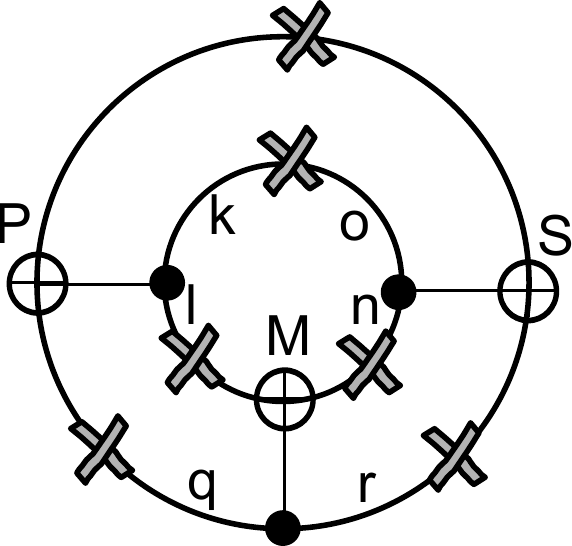}
}
\caption{Boolean variables assigned to wire segments for: a) CNOT $X$ transformations; b) CNOT $Z$ transformations; c) CNOT combined $X$ and $Z$ transformations (Sec.~\ref{sec:disc}); d) SWAP $X$ transformations; e) SWAP $Z$ transformations.}
\end{figure}

The functionality of a CNOT gate can be modelled by two Boolean expressions of the form Eq.~\ref{eq:braid}, because the transformations are of two types: $X$ and $Z$. In general, a wire segment is delimited by cut points or CNOT symbols ($\bullet$ or $\oplus$). In particular, Boolean variables denoted with small letters stand for variables representing wire segments ending at one of the symbols $\oplus$ or $\bullet$, and capitalised variables represent a wire segment running over one the CNOT symbols. To be more precise, in Fig.~\ref{fig:cnotxvars} $a$ and $b$ represent the wire segments having the target symbol $\oplus$ as an end point, and $C$ is the variable for the entire control wire. In Fig.~\ref{fig:cnotzvars} $a$ and $b$ represent the segments having $\bullet$ as an end point, and $C$ the entire target wire (contains $\oplus$).
\begin{align}
\mathcal{C}(a,b,C) = a \oplus b \oplus \lnot C \label{eq:braid}
\end{align}

The following example shows how Eq.~\ref{eq:braid} works and how to interpret the truth values of the variables. A definition of variables is offered in Section~\ref{sec:cuts}. If the control input of the CNOT is stabilised by $Z$, then $a \leftarrow true$ is replaced in Eq.~\ref{eq:braid} so that Eq.~\ref{eq:cxorb} results. The expression will be $true$ only if one of the variables is true; either $C = true$ or $b=true$. The first case corresponds to the result of multiplying Eq.~\ref{eq:iz} and ~\ref{eq:zi} ($Z_cZ_t \rightarrow I_cZ_t$), and the latter to Eq.~\ref{eq:zi}. Thus, a $true$ variable signals that a corresponding wire segment is stabilised, and a $false$ variable indicates the stabiliser $I$ (not stabilised). The possible stabiliser transformations of a CNOT are represented by each of the four clauses of the disjunctive normal form of Eq.~\ref{eq:braid}.
\begin{align}
\mathcal{C}(true, b, C) &= true \oplus b \oplus \lnot C = C \oplus b \label{eq:cxorb}
\end{align}

Expression~\ref{eq:cxorb} is a valid example of $X$ stabiliser transformations, too: if the target input (variable $a$) is stabilised by $X$, then $C=true$ corresponds to $X_cX_t \rightarrow X_cI_t$, and $b=true$ to Eq.~\ref{eq:ix}.

\subsection{Modelling Cuts}
\label{sec:cuts}

The CNOT gate discussion did not consider a circular representation because stabiliser transformations are functioning only in proper linear quantum circuits. This section introduces the Boolean modelling of cuts by the example of a circular single CNOT circuit (Fig.~\ref{circ:swapperm}). The only possible radial cut will result into two end points per wire: $c_{1,2}$ for the control wire, and $t_{1,2}$ for the target. Considering Fig.~\ref{fig:cnotxvars}, the variable $C$ is the segment spanned between $c_1$ and $c_2$, $a$ the segment between $t_1$ and $\oplus$, and $b$ the segment between $\oplus$ and $t_2$.

A linear quantum circuit is the result of a radial cut. In a circular representation with multiple CNOTs, this will generate two segment types: 1) segments delimited by a cut end point and a CNOT; 2) segments delimited by two distinct CNOTs. The first segment type represents wires reaching inputs or outputs, and the second type are circuit internal wires. However, the radial cut can be followed by additional cuts which affect only second type segments. Considering that a variable $s$ represented any of these segments and that, after a cut, the resulting subsegments are called $r$ and $t$, Eq.~\ref{eq:nrot} captures the Boolean behaviour before the cut: the Boolean variables are equivalent (the segments are joined), both can be either $true$ or false.
\begin{align}
\mathcal{J}(r,t) = \lnot r \oplus t \label{eq:nrot}
\end{align}

Cuts are modelled by not enforcing the subsegments to be equivalent, thus by not using expressions like Eq.~\ref{eq:nrot}. As a result, in the absence of cuts, segments delimited by two CNOTs are not considered independently, but as the result of joining the two subsegments generated after a potential cut. This observation leads to the Boolean variables interpretation (in the light of Section~\ref{sec:cnot}).

\begin{definition}
A Boolean variable represents a wire segment delimited by at least one cut end point.
\end{definition}

\begin{definition}
The truth value of a Boolean variable indicates if the qubit represented by the segment is stabilised or not.
\end{definition}



\subsection{Modelling an Entire Circular Circuit}

The Boolean model of an entire circular circuit includes, as mentioned in Section~\ref{sec:cnot}, two Boolean expressions ($\mathcal{B}_x$ and $\mathcal{B}_z$): one capturing $X$ and the other $Z$ stabiliser transformations. The expressions are built as \emph{conjunctions of clauses} (Eq.~\ref{eq:braid}, Eq.~\ref{eq:nrot}) formed after all the possible cut points were determined and the corresponding Boolean variables defined. The SWAP circular circuit is used once more as an example. Fig.~\ref{fig:swapxvars} and Fig.~\ref{fig:swapzvars} depict all the possible cuts, the resulting segments and the necessary variables for forming $\mathcal{B}_x$ and $\mathcal{B}_z$.
\begin{align}
\mathcal{B}_x & = \mathcal{C}(A, e,f)\mathcal{C}(G,b,c)\mathcal{C}(D,h,i)\nonumber\\
&\mathcal{J}(A,D)\mathcal{J}(A,b)\mathcal{J}(c,D)\mathcal{J}(e,i)\mathcal{J}(f,G)\mathcal{J}(G,h)\label{eq:swapx}\\
\mathcal{B}_z & = \mathcal{C}(P,k,l)\mathcal{C}(M,q,r)\mathcal{C}(S,n,o)\nonumber\\
&\mathcal{J}(k,o)\mathcal{J}(l,M)\mathcal{J}(M,n)\mathcal{J}(P,S)\mathcal{J}(P,q)\mathcal{J}(r,S)\label{eq:swapz}
\end{align}

Eq.~\ref{eq:swapx} and ~\ref{eq:swapz} model the circular SWAP, where no cuts were made. In order to generate a functioning SWAP circuit, the necessary radial cut will remove the clauses $\mathcal{J}(A,D)$ and $\mathcal{J}(e,i)$ from $\mathcal{B}_x$, and the clauses $\mathcal{J}(k,o)$ and $\mathcal{J}(P,S)$ from $\mathcal{B}_z$. The Boolean expressions resulting after the removals will represent the circuit in Fig.~\ref{circ:swap}.

In order to generate the circular permutation of the SWAP and to obtain the circuit that implements a single CNOT (Fig.~\ref{circ:swapperm}) the radial cut could remove $\mathcal{J}(c,D)$ and $\mathcal{J}(G,h)$ from $\mathcal{B}_x$, and $\mathcal{J}(M,n)$ and $\mathcal{J}(r,S)$ from $\mathcal{B}_z$. The teleported CNOT circuit (Fig.~\ref{circ:rcnot}) is the result of performing the cuts $\mathcal{J}(A,D)$; $\mathcal{J}(e,i)$; $\mathcal{J}(A,b)$; $\mathcal{J}(c,D)$ in $\mathcal{B}_x$, and the cuts $\mathcal{J}(k,o)$; $\mathcal{J}(P,S)$; $\mathcal{J}(l,M)$; $\mathcal{J}(M,n)$ in $\mathcal{B}_z$. Finally, the selective destination teleportation circuit (Fig.~\ref{circ:seldest}) is obtained by cutting $\mathcal{J}(c,D)$; $\mathcal{J}(G,h)$; $\mathcal{J}(A,D)$; $\mathcal{J}(f,G)$ in $\mathcal{B}_x$ and $\mathcal{J}(M,n)$; $\mathcal{J}(r,S)$; $\mathcal{J}(k,o)$; $\mathcal{J}(P,q)$ in $\mathcal{B}_z$.


\subsection{Discussion}
\label{sec:disc}
Boolean models of circular CNOT circuits include two expressions, and this structure was chosen because each expression is equivalent to a linear equations system: each clause is a linear equation (XOR is a linear function). The equivalence between the Boolean model and a stabiliser table obtained after simulating a stabiliser circuit can be observed, too. Stabiliser table operations are performed as if the table were a linear equations system (e.g. Gaussian elimination is used for determining individual qubit measurement results) \cite{aaronson2004improved}. A second argument for the chosen structure was that in a CNOT circuit the X and the Z stabilisers transformations do not interact one with another. This would have not been the case if, for example, Hadamard ($H=PVP$) gates were included in the circuit. The Hadamard transforms the input $X$ stabiliser into $Z$, and vice versa. Similarly, if the circular circuits had included CPHASE gates, X and Z would have been referenced in the same stabiliser transformations.

The manner in which cuts and Boolean variables were defined could have been simplified if a single Boolean expression per CNOT had modelled both the X and Z stabiliser transformations. In this situation, a wire segment is determined by exactly one cut point and a CNOT element ($\bullet$ or $\oplus$). Each wire segment has a Boolean variable attached, and for a single CNOT circuit the segments and the variables are similar to Fig.~\ref{fig:cnotvars}, and Eq.~\ref{eq:all} models all the stabiliser transformations.
\begin{align}
F(a,b,c,d) = x\mathcal{C}(a,c,d)(a \oplus \lnot b) \oplus (\lnot x)\mathcal{C}(c,a,b)(c \oplus \lnot d) \label{eq:all}
\end{align}

The previous expression references the function defined in Eq.~\ref{eq:braid} and introduces two additional variables $x$ and $z$. If one would like to compute the $X$ transformation of a CNOT the $x$ variable needs to be set to true, and for the $Z$ transformation the variable has to be $false$. The complete Boolean model of a circular CNOT circuit results after conjugating for all the CNOTs the corresponding expressions of form Eq.~\ref{eq:all}. The Boolean model of the cuts will remain the same.

Irrespective of the used model (with a single or two Boolean expressions), the temporal ordering of the gates is not relevant. The traversal direction of the gates is not important when trying to determine a stabiliser transformation computed by the modelled circuit. The CNOT gate is reversible, its Boolean model captures its reversibility. If a variable is set to $true$ and the truth value of another variable has to be computed, the direction of the stabiliser transformations (equivalent to gate traversal order) is dictated by the modelled Boolean constraints.

It can be also noted that the Boolean expressions capture the CNOT commutativity inside the circuit (Fig.~\ref{circ:cnotcomm}). This is due to how the segments were defined: in $\mathcal{B}_x$ the capitalised variables represent segments containing the $\bullet$, and in $\mathcal{B}_z$ the capitalised variables stand for segments containing $\oplus$. For neighbouring CNOTs having the control ($\mathcal{B}_x$) or the target ($\mathcal{B}_z$) on the same qubit, the capitalised variables need to be interchanged in order to commute the gates. Variables of joined (uncut) segments can be interchanged due to Eq.~\ref{eq:nrot}, which is the Boolean equivalence relation between two variables.

\section{ICM Circuits are Instances of Circular CNOT Circuits}
\label{sec:constr}

There are two strategies for constructing a quantum circuit from a circular representation. The first is to make a single radial cut, and the second is to make additional single cuts following a radial cut. A radial cut generates an ICM circuit having an equal number of wires to the circular representation, while each additional single cut introduces an additional qubit (wire) in the circuit. This is observed after comparing Fig.~\ref{fig:swap0} and \ref{fig:swap1}. Consequently, circuits obtained after radial cuts have indeed cyclic permuted gate lists. The second construction strategy, however, does not preserve the number of wires from the circular representation, and the resulting gate lists are cyclic permutations only in the sense of the CNOT ordering and direction (the affected qubits are not identical).

The position of the cuts dictates the chosen gate list permutation of the resulting ICM circuit, but the circuit will not implement any function until its qubits are configured. Configuration is the process of selecting qubit initialisation and measurement basis. In general, a quantum circuit includes both input/output and ancillae qubits (have predetermined initialisation and measurement basis). In particular, for ICM circuits the basis determine either the rotational gate being implemented or supplemental decisions required during information processing. An example for the latter situation offers the selective destination teleportation circuit which acts like a multiplexer: the third or the fourth qubit outputs the state of one the first qubit depending on the measurement basis of these first two qubits (Fig.~\ref{circ:icmseldest}). Non-ancillae qubits take the states supplied to the circuit (inputs) or are used for reading out states after circuit execution (outputs). Each end point introduced after a cut will represent either an ancilla or an input/output qubit. Thus, the construction of ICM circuits from circular CNOT circuits requires three steps: 1) correctly cut and choose traversal direction of the circular circuit; 2) select which end points belong to ancillae and which not; 3) choose the initialisation and measurement basis of the ancillae. The example of the circular SWAP circuit in the previous Section illustrates these steps.

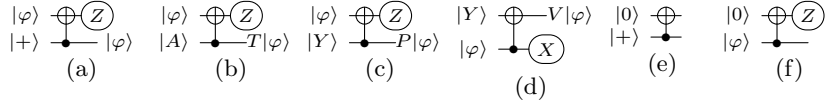
\begin{figure}[t!]
\scriptsize
\centering
\subfloat[ ]{
\Qcircuit @C=.3em @R=.5em {
		\lstick{\ket{\varphi}}& \targ & \measure{Z} \\
		\lstick{\ket{+}} & \ctrl{-1} & \qw & \ket{\varphi}
	}
}
\hfil
\subfloat[ ]{
\label{circ:tgate}
\Qcircuit @C=.3em @R=.5em {
		\lstick{\ket{\varphi}}& \targ & \measure{Z}  \\
		\lstick{\ket{A}} & \ctrl{-1} & \qw & T\ket{\varphi} \\
	}
}
\hfil
\subfloat[ ]{
\label{circ:pgate}
\Qcircuit @C=.3em @R=.5em {
		\lstick{\ket{\varphi}}& \targ & \measure{Z}\\
		\lstick{\ket{Y}} & \ctrl{-1} & \qw & P\ket{\varphi}
	}
}
\hfil
\subfloat[ ]{
\label{circ:vgate}
\Qcircuit @C=.3em @R=.5em {
		\lstick{\ket{Y}}& \targ & \qw & V\ket{\varphi}\\
		\lstick{\ket{\varphi}} & \ctrl{-1} & \measure{X}
	}
}
\hfil
\subfloat[ ]{
\Qcircuit @C=.3em @R=.5em {
		\lstick{\ket{0}}& \targ & \qw \\
		\lstick{\ket{+}} & \ctrl{-1} & \qw
	}
}
\hfil
\subfloat[ ]{
\Qcircuit @C=.3em @R=.5em {
		\lstick{\ket{0}}& \targ & \measure{Z} \\
		\lstick{\ket{\varphi}} & \ctrl{-1} & \qw
	}
}
\caption{The ICM circuits have the same CNOT gate structure, but different qubit initialisation and measurements: a) information teleportation circuit; b) teleported implementation of the T gate; c) teleported implementation of the P gate; d) teleported implementation of the V gate; e) construction of a Bell pair; f) measurement of the $Z$ operator. The qubits marked with $\ket{\varphi}$ are input/output, and all the others are ancillae. All the circuits will have the same circular CNOT circuit representation, thus the same Boolean model.}
\label{circ:icmversions}
\end{figure}

There are two abstraction levels, conforming to the previously listed steps, necessary for highlighting circular CNOT circuit capabilities. The \emph{first level} is represented by circuits having the same CNOT structure but different initialisation/measurement basis (e.g. Fig.~\ref{circ:icmversions}). At this level all the circuits implement the same underlying stabiliser transformations, because the CNOTs are arranged identically. The \emph{second level} is the circular CNOT representation and its Boolean model, which abstracts all the circuits that have the same set of gates, but arranged as cyclic permutations. In contrast to the first level, at the second level the abstracted circuits do not implement the same stabiliser transformations, because their gates are arranged differently (once more compare Fig.~\ref{circ:swap} and~\ref{circ:swapperm}). As a result, constructing an ICM circuit from a circular representation is equivalent to selecting an ICM circuit instance from the set of abstracted circuits. It is straightforward to compute the circuit's stabiliser transformations using the Boolean expressions resulting after the cuts.

The main advantage of circular CNOT circuits is that they abstract a large set of ICM circuits, and by this their Boolean model is the abstraction of multiple related possible stabiliser transformations. Each different cut choice in the circular representation has the potential to result in a different ICM circuit structure with correspondingly different stabiliser transformations. It is beneficial to have the possibility to generate/select a specific set of stabiliser transformations which are required for a particular quantum computation, because scalable error corrected quantum circuits are equivalent to ICM circuits. Although including only CNOTs, their computational universality is given by the appropriate initialisation and measurement basis, and it is advantageous to derive sets of related quantum circuits and understand their structure.

As a conclusion, a circular CNOT circuit can be formed as a generalisation for any fault-tolerant error corrected circuit.

\section{Example: The ICM Toffoli Gate}

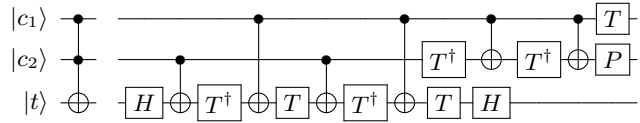
\begin{figure}
\centerline{
\small{
\Qcircuit @C=.3em @R=.3em {
	\lstick{\ket{c_1}}&\ctrl{1}&\qw& &&& \qw & \qw &\qw &\ctrl{2} &\qw&\qw&\qw&\ctrl{2} &\qw&\ctrl{1}&\qw&\ctrl{1}&\gate{T} &\qw\\
	\lstick{\ket{c_2}}&\ctrl{1}&\qw& &&& \qw&\ctrl{1}&\qw&\qw&\qw&\ctrl{1}&\qw&\qw&\gate{T^\dagger}&\targ&\gate{T^\dagger}&\targ&\gate{P}&\qw\\
	\lstick{\ket{t}}&\targ&\qw& &&& \gate{H} & \targ & \gate{T^\dagger} & \targ & \gate{T} & \targ & \gate{T^\dagger} & \targ & \gate{T} & \gate{H} &\qw&\qw&\qw&\qw
	}
}
}
\caption{Toffoli gate using CNOT, $T$ ($T^\dagger$), $P$ and Hadamard gates \cite{NC00}.}
\label{circ:toffoli}
\end{figure}

\begin{figure}[t!]
\includegraphics[width=0.9\columnwidth]{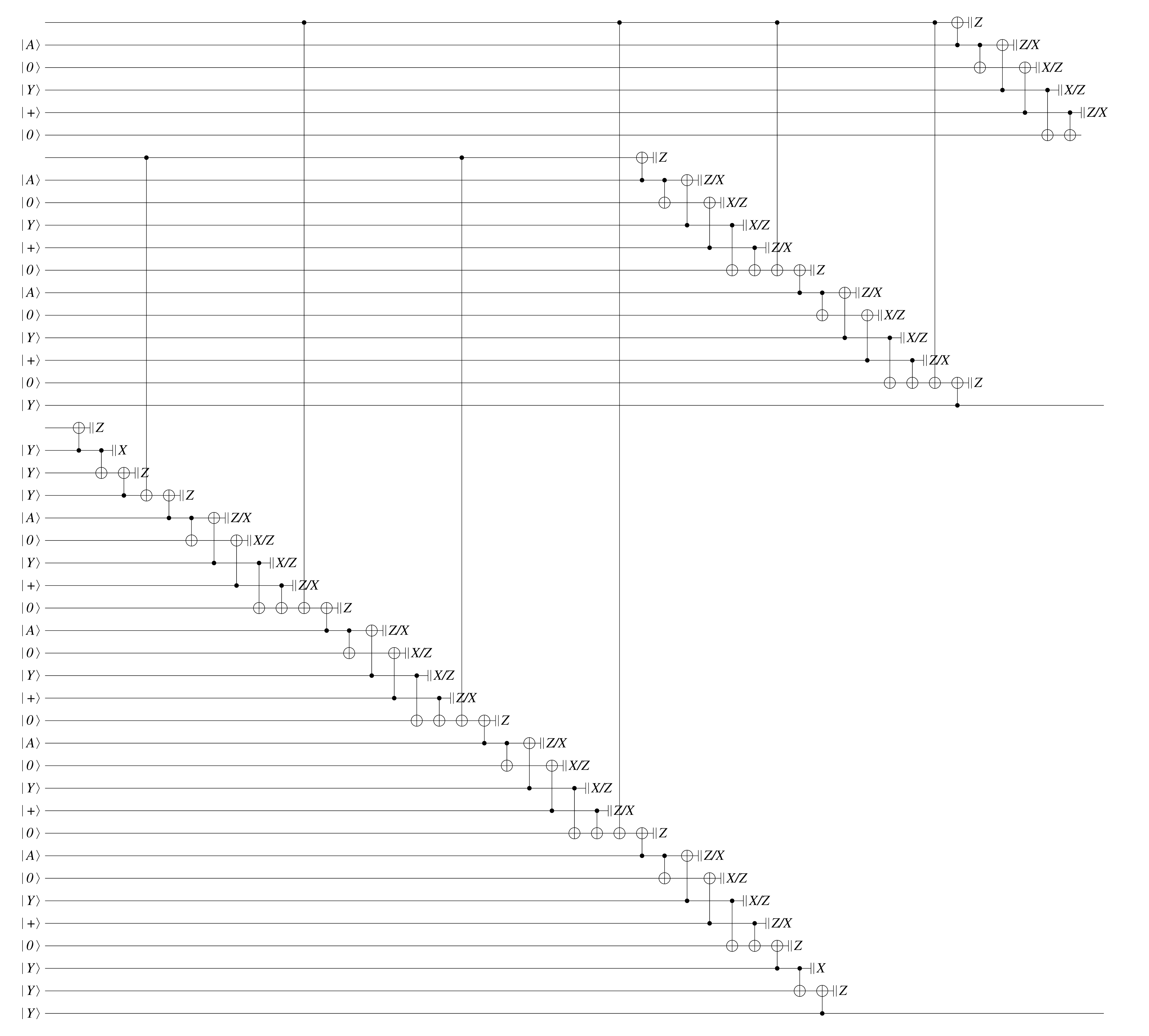}
\caption{The ICM Toffoli gate implementation. Additional qubits are introduced because each of the $T$ and Hadamard gates from Fig.~\ref{circ:toffoli} is implemented using the teleported rotational gates from Fig.~\ref{circ:tgate},\ref{circ:pgate},\ref{circ:vgate} and with the use of measurement-controlled teleportation subcircuits (e.g. Fig.~\ref{circ:icmseldest}). The configurable measurement basis ($Z/X$ and $X/Z$) are an ICM mechanism for controlling the information flow in the circuit.}
\label{fig:icmtoffoli}
\end{figure}

Reversible circuits make extensive use of the Toffoli gate because it is classically universal (can simulate the classical AND, OR and NOT gates). Quantum computing architectures, especially large-scale error corrected ones, do not support the direct application of this gate.  Therefore, the Toffoli gate needs to be firstly decomposed into a sequence of architecture specific gates. The decomposition into $T$ and Hadamard gates, and the ICM implementation of the Toffoli gate are presented in Fig.~\ref{circ:toffoli} and~\ref{fig:icmtoffoli}.

\begin{figure}[t!]
\includegraphics[width=\columnwidth]{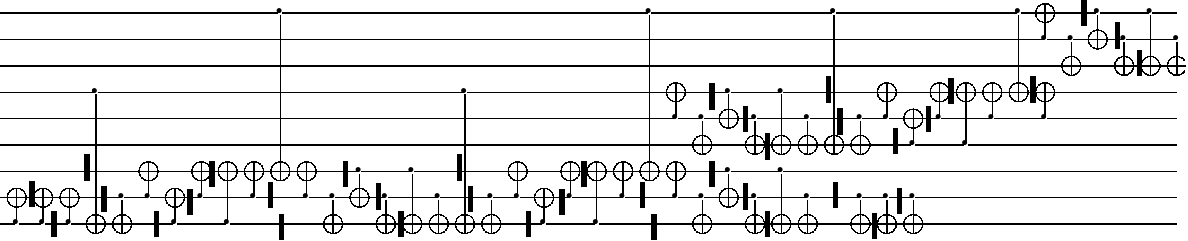}
\caption{The circular circuit of the ICM Toffoli decomposition. The circular representation is obtained after joining the left and right wire end points. The linear representation simplifies the visualisation. The cuts necessary to reconstruct the ICM equivalent circuit (Fig.~\ref{fig:icmtoffoli}) are depicted with black horizontal bars.}
\label{fig:cyclictoffoli}
\end{figure}

The previous sections discussed the construction of ICM circuits from the circular representation, but this section will backtrack the steps from ICM to circular CNOT circuit (Fig.~\ref{fig:cyclictoffoli}). Firstly, for the ICM Toffoli gate implementation, the (I)nitialisations and the (M)easurement components are removed (backtrack second and third steps from Sec.~\ref{sec:constr}). Secondly, all qubits operated by a single CNOT are uncut (joined). At this stage circuits like Fig.~\ref{circ:rcnot} are backtracked to a structure like Fig.~\ref{circ:swap}. Thirdly, all the remaining wire end points are looped, such that a circular structure finally emerges. The circular CNOT circuit of the ICM Toffoli is depicted in Fig.~\ref{fig:cyclictoffoli}. Algorithm~\ref{alg:1} summarises the circular CNOT construction using pseudo code. The algorithm assumes that circuit inputs are on the right and outputs on the left and that each CNOT is applied at a specific time $t$. The construction starts with the bottom most qubit. For the current qubit to be processed, the algorithm searches for the first CNOT gate that is applied on it (e.g. at time $min$), and selects the closest upper qubit which is not affected by a CNOT applied at time $\ge min$. The current and the closest upper wire are joined.

Comparing Fig.~\ref{fig:cyclictoffoli} and~\ref{circ:toffoli} against Fig.~\ref{fig:icmtoffoli} shows that the circular representation uses less wires than the ICM equivalent (9 vs. 45), and increases the number of wires of the non-ICM decomposition by a factor of three (9 vs. 3). The cause of this is that single CNOT operated qubits are uncut. The circular representation shows that potential future ICM circuit optimisation techniques should consider \emph{qubit reuse} techniques.

\begin{algorithm}[t]
\caption{Construction of Circular CNOT Circuit}
\label{alg:1}
\begin{algorithmic}[1]
\REQUIRE{$icm$ an ICM circuit}
\STATE{$nrq \leftarrow icm.qubits$}
\FORALL{$qub \in [nrq, 1]$}
	\STATE{$min$ time of left-most (first) $\bullet$ or $\oplus$ on wire $qub$}
	\STATE{$pqub$ first wire so that: 1) $pqub < nrq$, and 2) $pqub$ is not used by any CNOT with time $\ge min$}
	\STATE{Join left end point of $qub$ (input) with right end point of $pqub$ (output)}
\ENDFOR
\end{algorithmic}
\end{algorithm}


\section{Applications of Circular CNOT Circuits}

The stabiliser transformations supported by a circular CNOT circuit are representative for an entire set of ICM circuits. The number of generated ICM circuits is a function of the number of cuts allowed on the circular wires. However, the number of equivalent generated circuits is not known for the moment. Future work on circular CNOT circuits will evaluate the number of equivalent circuits, but also on using these for the test and verification of ICM circuits.

The circular CNOT circuit representation can be used to model single missing gate faults (SMGF) when testing ICM quantum circuits. An SMGF is defined as a missing gate from the ideal gate list of the circuit under test. Such faults are detected by applying appropriate tests (initialising qubits according to a pattern) at circuit inputs and reading out the computed states at circuit output \cite{polian2005family}. Methods for determining appropriate tests were investigated for example in \cite{patel2004fault,polian2005family}. Because ICM circuit gate lists include only CNOTs, a CNOT SMGF is equivalent to having a control \emph{stuck at} $\ket{0}$; thus, the target is never affected. Considering a set of cuts that generate the tested circuit, the fault is modelled by introducing at most two additional cuts around the control of the CNOT, so that an ancilla qubit results (similar to Fig.~\ref{circ:rcnot}). The ancilla will have its state stuck at $\ket{0}$.

\begin{figure}[t!]
\centering
\subfloat[ ]{
	\includegraphics[width=0.25\columnwidth]{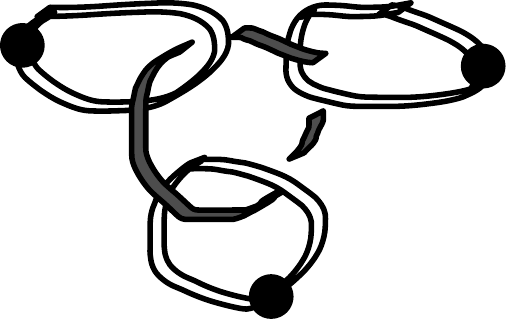}
}
\hfil
\subfloat[ ]{
	\includegraphics[width=0.25\columnwidth]{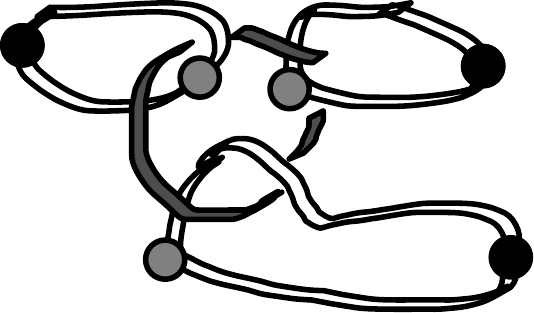}
}
\hfil
\subfloat[ ]{
	\includegraphics[width=0.25\columnwidth]{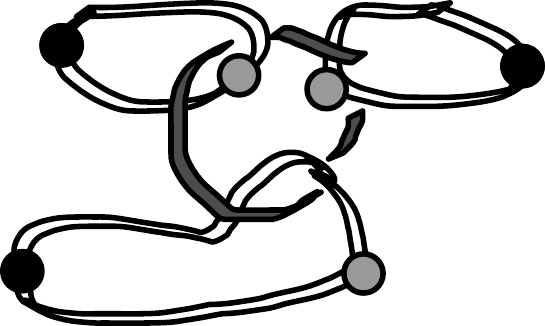}
}
\caption{Four qubits are encoded as strands and braided. The figures contain three braids (CNOTs), as the grey strand is braided with the three white strands. The black points denote potential input/output locations. The grey points represent additionally included input/outputs. The strands can be arbitrarily deformed, so that b) and c) have a CNOT structure equivalent to a). The figures b) and c) assume that there is a horizontal temporal axis, and that inputs are on the left while outputs on the right side. The functionality of the ICM equivalent circuits depends on the initialisation and measurement basis chosen for the qubits.}
\label{fig:strands}
\end{figure}

The verification of ICM circuits is a problem encountered in the context of fault-tolerant quantum computations. Large scale circuits need to be error corrected in order to achieve a certain fault-tolerance threshold, and one of the most promising error correction techniques is based on topological properties of the encoded information \cite{FMM13}. In that particular computational model information is encoded as strands and braids are the implementation of CNOT gates, thus the resulting circuits have an ICM interpretation. The information strands can be arbitrarily deformed as long as the braiding structure is left unchanged and, furthermore, circuit inputs and outputs can be placed anywhere on a strand. The placement is not guaranteed to make any computational sense, but it does not invalidate the strands (encoded qubit states) or the braids (CNOTs). The main issue with such error corrected circuits is that their ICM interpretation is a function of input/output location: the same circuit description in terms of strands and braids can be interpreted as different ICM circuit. It can be seen (e.g. Fig.~\ref{fig:strands}) that input/output placement on strands is similar to cutting a circular representation. For this reason, the proposed circular representation is a valuable tool for verifying topologically error corrected ICM circuits: which cuts need to be made, and which traversal direction is required for the resulting ICM circuit to support a given set of stabiliser transformations? The support guarantees that the structure of the circuit (number of qubits and CNOT gate list) is correct, and that if the circuit were configured with corresponding qubit initialisations and measurements, a correct sequence of teleported rotational gates (T,P,V) and CNOTs would be implemented.

\section{Conclusion}

This work introduced circular CNOT circuits and their Boolean model. Two Boolean expressions are capturing all the possible stabiliser transformations supported by the circular circuits, one for $X$ stabiliser transformations and another for $Z$ transformations. Circular circuits can be transformed into fault-tolerant error corrected quantum circuits after performing a well defined set of cuts. The resulting circuits are the basis of ICM circuits, which are required for universal scalable fault-tolerant quantum computing. ICM circuits consist entirely of qubit initialisations, CNOTs (because all the single qubit quantum gates are implemented by teleportation) and qubit measurements. ICM circuits originating from the same circular CNOT circuit will have gate lists which are cyclic permutations of one another. Having modelled all the stabiliser mappings supported by a circular circuit, it is straightforward to infer the stabiliser transformations of a particular ICM instance.

Applications of circular CNOT circuits were enumerated in conjunction with their ICM transformation and showcase new possibilities for the design of quantum circuits. Future work will detail circular CNOT circuit based methods for optimisation, SMGF testing and verification of ICM circuits.

\bibliographystyle{plain}
\bibliography{circular}

\end{document}